\documentclass[%
 aip,
cp, 
 amsmath,amssymb,
 reprint,%
]{revtex4-2}

\usepackage{graphicx}
\usepackage{dcolumn}
\usepackage{bm}
\usepackage{hyperref}
\hypersetup{colorlinks=true, allcolors=blue}
\usepackage[utf8]{inputenc}
\usepackage[T1]{fontenc}
\usepackage{mathptmx} 
\usepackage{xcolor}

\begin{document}
\title{Mach Number Dependence of Flow Instability around a Spiked Body}

\author{Ashish Vashishtha} 
 \email[Corresponding author: ]{vashish.aero@gmail.com}
\affiliation{Joint First Author}
\affiliation{
  Assistant Lecturer,
  Institute of Technology Carlow, R93 V960 IRELAND
}

\author{Shashank Khurana}
\email[]{skhurana@dubai.bits-pilani.ac.in}
\affiliation{Joint First Author}
\affiliation{%
Assistant Professor, Department of Mechanical Engineering, BITS Pilani Dubai Campus, 345055, UAE
}%

\date{\today} 

\begin{abstract}
A forward-facing aerospike have been identified as a passive flow control device for enhancing the aerodynamic efficiency and reducing the heat transfer in high-speed flows. In addition, it has been reported that the presence of a spike brings in unsteadiness in the form of oscillation and pulsation to the structure. Previous researchers have investigated the aerothermodynamic coefficients, together with offering a detailed explanation of the flow physics and associated unsteadiness, and their dependence on the spike’s geometric characteristics (spike nose, and length-to-fore-body diameter ratio, $L/D$). This work \textcolor{black}{focuses} on ascertaining the role of flow speeds (free-stream Mach number), and their energy content, in governing the physics around a spiked body, which is yet to be established. Numerical investigation has been carried out using axisymmetric Navier-Stokes laminar flow solver for Mach number range of 2.0 to 7.0. A round-tip spike with flat-face cylindrical after-body have been simulated for spike length ratio of $L/D$ = 2.0, with spike diameter to fore-body diameter of 0.1. The flow unsteadiness has been analyzed with drag and pressure coefficients variation at different Mach numbers. It was found that the flow field around the spiked blunt nose behaves in pulsation mode at lower Mach numbers 2, 3 and transition to oscillatory mode at higher Mach numbers 5, 6 and 7, while remain almost stable at Mach 4. \textcolor{black}{The limit of Strouhal Number for characterizing the pulsation and oscillation modes at various Mach numbers for spike length of $L/D = 2$ with flat after-body is observed as 0.2, however it may very well depend on other geometric parameters of spike and after-body}.
\end{abstract}

\maketitle

\section{\label{sec:level1}Introduction\protect}
An object flying at supersonic and hypersonic speeds is subjected to high drag because of formation of shock wave in front of its nose. The strength of shock wave governs the drag experienced by the object and producing of aerodynamic heating at high supersonic and hypersonic flows. The manipulation of this shock wave, aiming to control its strength was motivation of various flow control studies \cite{ahmed2020}. However, utilization of various flow control methods has not been realized or effectively utilized because of either unsteadiness associated with the control or complication of systems. The flow control methods for shock manipulation can be classified into active and passive controls. The mechanical spike mounted in front of a blunt body, has been deemed beneficial as a passive flow control technique towards reducing the heat transfer and aerodynamic drag in supersonic and hypersonic flows \cite{ahmed2020}. The flow unsteadiness in the form of pulsation and oscillation associated with spike has been widely researched and studied in the past \cite{maull, wood, Kenworthy}. %
A hemispherical cavity in front of hypersonic flow is also considered an effective flow control to increase aerodynamic drag and reduce heating at hypersonic flows, but it is also subjected to very highly unsteady flow-field, resulting in uncontrolled lift and side forces \cite{utokyo1, utokyo2}. Few passive control methods have been suggested for frontal cavity and found effective in controlling the unsteadiness \cite{utokyo3}. In active flow control methods, the counter-jet flow \cite{ahmed2020} has also been found effective method, creating flow spike to push the shock wave away from the blunt nose, but require additional system to inject high-pressure gas into the stagnation zone. And, it may also be associated with unsteadiness, depends on the injected pressure and gas, exhibits short and long penetration mode. Further, energy addition by electric discharge or laser was also found useful, but again require additional complicated system and repetitive energy addition may lead to unsteady flow field. Hybrid active control method such as direct hydrogen injection \cite{vashishtha2021} was also investigated, which creates gas spike as well as add energy by self-ignition of hydrogen, recently. However, the proposed system was effective than air injection, but associated with high unsteady flow-field and standing flame. Irrespective of the fact that there is an active or a passive flow controls method to manipulate the shock wave, all inherit unsteadiness because of geometric characteristics of control or operative conditions. This study focuses on unsteady behaviour of mechanical spike in high-speed flows. Festzy et al. \cite{festzy1, festzy2, festzy3} have studied the unsteadiness related to mechanical spike and provided explanation of the same at supersonic and hypersonic regimes. Recently, Vashishtha and Khurana \cite{shashank1,shashank2} assessed the unsteady flow characteristics associated with a spiked body, and their dependence on geometric parameters of spike (including spike length and spike nose tip), and the attached afterbody of different shapes. In addition to the geometric parameters, it is believed that from the perspective of practical applications, the flow speeds (corresponding to the local atmospheric conditions in flight for a aerospike vehicle) could play an important role in the frequency and amplitude of pulsations generated. A thorough literature survey reveals the absence of such a research. Hence, the numerical investigation has been performed in this study for single spike length $L/D = 2.0$ at a constant altitude of 30 km, with varying free-stream Mach numbers 2-7, ranging from supersonic to hypersonic flow regime. The study is motivated to develop understanding and possibly offer an explanation on the dependence of flow Mach numbers on the aerodynamic force coefficients, together with computing the magnitude of pulsations. 
\section{\label{sec:level2}Numerical Method}
A two-dimensional structured grids have been generated in the computational domain for a hemispherical tipped spike of length $L = 2D$ with a flat-face cylindrical after-body. The unsteady axi-symmetric Navier-Stokes equations for laminar compressible flows have been solved here. The spatial inviscid fluxes are evaluated by Lious all-speed AUSM (Advection Upstream Splitting Method) + up scheme \cite{liou} with upwind biased third order MUSCL (monotonic upstream-centered scheme for conservative laws) interpolation, while viscous fluxes as well as source terms (because of axi-symmetry) are evaluated by using second order central difference scheme. The third order TVD Runge-Kutta Method \cite{shu} was employed for time integration. \textcolor{black}{Within the scope of this study, in order to analyze the flow instabilities with flow features and drag coefficient, the choice of two-dimensional axi-symmetric solver is adequate based on previous numerical studies \citep{festzy1,festzy2,festzy3,vashishtha2021}} 
\begin{figure}[!b]
\centering
\includegraphics[scale= 0.8]{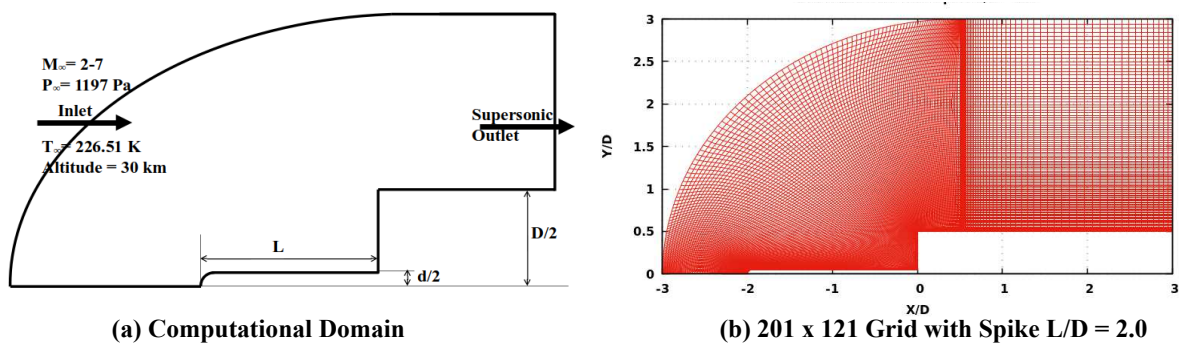}
\caption{\label{Fig01} (a) Configuration of Spiked Body (b) Computation Domain and Grid.}
\end{figure}
\begin{table}[!b]
\caption{\label{table:table1} Free-stream Conditions}
\begin{tabular}{p{5cm}p{4cm}} \hline \hline
Property & Specification \\ \hline
Mach Number ($M_{\infty}$) & 2.0 - 7.0\\
Altitude & 30 km. \\
Free-stream Pressure ($P_{\infty}$) & 1197 Pa \\ 
Free-stream Temperature ($T_{\infty}$) & 226.51 K \\
Free-stream Speed ($U_{\infty}$) & 603.3 - 2117.7 m/s \\
Reference Length & 40 mm \\
Reynolds Number & (0.03 - 0.105) $\times$ $10^6$ \\ \hline \hline
\end{tabular}
\end{table}
Figure \ref{Fig01}a shows the overall computation domain, consists of hemispherical-spiked body for $L/D = 2.0$ in front of flat-face of blunt nose after-body. Here, $L$ is the spike length and $D$ is the maximum cross-sectional diameter of the base blunt nose (in current study, $D = 40 mm$). The spike diameter (d) has been considered as $d/D = 0.1$. Figure 1(b) shows the structured grid generated using elliptic grid generation method. In this study, the grid size of $201 \times 121$ has been used for all the cases as shown in Fig. \ref{Fig01}b. The grid size is adopted based on the previous grid independence studies \cite{shashank1, shashank2} performed. The minimum non-dimensional grid spacing near the wall is used here as $1 \times 10^{-4}$. The inlet boundary conditions are considered for the flow conditions corresponding to Mach numbers of 2-7 at an altitude of 30 km, such that the Reynolds number varies between 0.03 – 0.105 million. The reference length or diameter of blunt nose have been used as 40 mm for current study. Table \ref{table:table1} summarizes the free-stream inlet boundary conditions, that are simulated in this study. The outlet is treated as supersonic outlet with first order interpolation of fluxes from the inner domain. At the wall, no-slip boundary condition along with constant wall temperature of 300 K is assumed. The flow-field is initiated with low velocity around the spike and blunt nose and is considered to have an impulse start. All the computations have used physical time-step of $1 \times 10^{-9}$ s and have been performed for a duration of 4 ms, which leads to CFL variation from 0.16 - 0.6 for explicit time integration. The overall drag coefficient ($C_D$)and point pressure coefficient at the flat face of after-body (at $D/4$ distance from corner, called as $P_{D/2}$ have been used to analyze the unsteadiness along with contour plots of Mach number.%
\section{Results and Discussions}
All the simulations have been performed for only one spike length $L/D = 2$ and one spike diameter ($d/D = 0.1$). The flow conditions has been selected based on earth's altitude of 30 km. \textcolor{black}{At this selected altitude, the environment pressure and temperature can cause low-Reynolds number flow, where laminar Navier-Strokes solver may work very well as well as practical flight altitude of supersonic and hypersonic flight}. The Mach number ($M_{\infty}$) was varied between 2 - 7 by only changing the speed, while the free-stream pressure and temperature remains same for all the simulated cases. The flow regime covers theoretically supersonic (Mach 2, 3 and 4) and hypersonic (Mach 5, 6 and 7) flows. The results are discussed by plotting time dependent drag coefficient as well as non-dimensional pressure ($P/P_{\infty}$) at front face location with quarter diameter height from the center-line. The free-stream pressure ($P_{\infty}$ is same for all the simulated cases, hence it has been used instead of dynamic pressure ($(\gamma P_{\infty}M_{\infty}^{2}/2$, where $\gamma$ is specific heat ratio of air) to non-dimensionalize the computed pressure. Further, Mach contours have been used to explain the unsteady behavior at various free-stream Mach numbers.%
\subsection{Drag and Pressure Variations}
\begin{figure}[b]
\centering
\includegraphics[scale= 0.8]{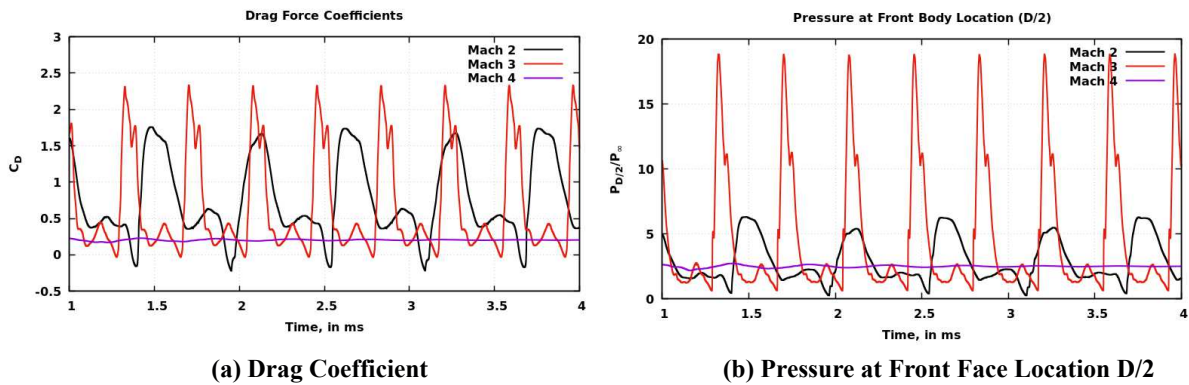}
\caption{\label{Fig02} (a) Drag Coefficient (b) Non-dimensional Pressure at $D/2$ location in supersonic range.}
\end{figure}

\begin{figure}[!htb]
\centering
\includegraphics[scale= 0.8]{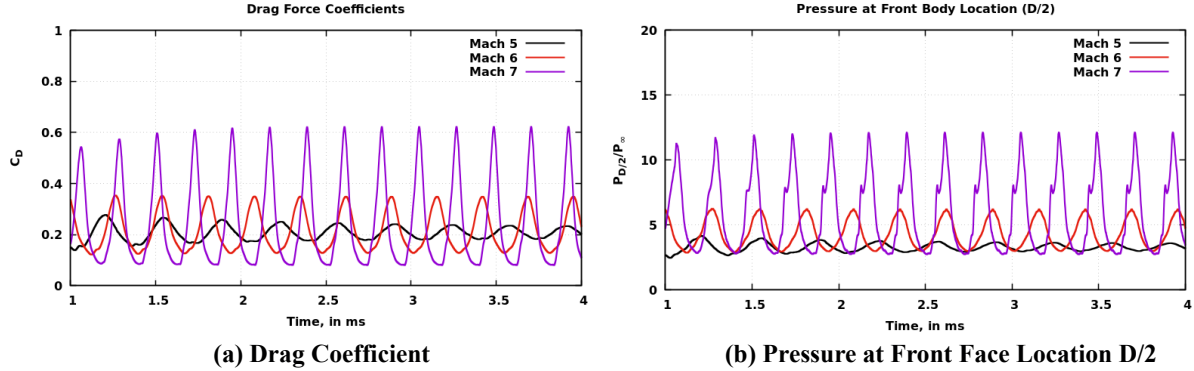}
\caption{\label{Fig03} (a) Drag Coefficient (b) Non-dimensional Pressure at D/2 location in hypersonic range.}
\end{figure}

The drag coefficient \textcolor{black}{and pressure data at quarter diameter height have been recorded at each physical time step of  $1 \mu s$}. Both are plotted in Figs. \ref{Fig02} and \ref{Fig03} for supersonic and hypersonic Mach numbers, respectively. These figures show the plots for the duration of 1 to 4 ms. As the simulations are started from impulse start, initial 1 ms data is not plotted here. It is evident from Fig. \ref{Fig02}a that flow remains highly unsteady at Mach 2 and 3, while at Mach 4 the flow-field becomes stable and oscillates by very small amplitude. The amplitude of unsteadiness increases from Mach 2 to 3, which reflects that the bow-shock envelop may become closer to the corner and shows high unsteadiness because of shock, \textcolor{black}{shear layer and re-circulation interaction with corner}. The pressure plots in Fig. \ref{Fig02}b also depicts the same. 
The amplitude in pressure plots provide absolute picture of pressure variations as it is non-dimensionalized by same free-stream pressure for all the Mach number cases. In the drag coefficient, it may not provide the absolute picture as drag force is non-dimensionalized by dynamic pressure ($\gamma P_{\infty}M_{\infty}^{2}/2$), which increases with increase in Mach number, \textcolor{black}{while the free-stream pressure remains constant}. Figure \ref{Fig02}a shows that the shock may remain closer to the body at Mach 3 and exerts higher pressure at the front face. The frequency of high peaks are also higher at Mach 3 in comparison to Mach 2. Figure 3 shows the drag coefficients and non-dimensional pressure variation for hypersonic Mach numbers 5, 6 and 7. It should be noted that Mach 5 is not a hard boundary between supersonic and hypersonic flow, hence it is said here as theoretically hypersonic flow regime. The fluctuations in drag coefficient are smooth and the amplitude of fluctuations increases with increase in Mach number as well as number of high peaks in given time. The pressure fluctuations \textcolor{black}{plot} also shows similar trend to drag coefficient, however the amplitude of pressure fluctuations increases more than double with unit increase in Mach number. The fluctuations in supersonic and hypersonic range contrast as supersonic flow regime fluctuation in drag coefficients and pressure ratio shows multiple local peaks and troughs in a single cycle, while in hypersonic regime it is slightly smoother. It can be understood that the incoming momentum increases with increase in Mach number at same altitude, which can lead to modify the bow shock envelop closer to the body. The interaction of bow shock with the after-body will define the unsteady nature of flow-field in front of blunt nose. Hence, in the next section Mach contours are plotted at different Mach number and have been analysed to understand the mechanism of flow unsteadiness.%
\subsection{Mach Contours}
In this section, the nature of bow shock unsteadiness has been discussed with Mach contour  \textcolor{black}{plots} at various Mach numbers range from 2 - 7. It is understood from the previous studies by Festzy et al. \cite{festzy2, festzy3} that the bow shock in front of spiked body can exhibit high amplitude pulsation or small amplitude oscillation modes which  \textcolor{black}{may depend} on various flow parameters. The pulsation mode consists of three main processes during the one cycle as: inflation, withhold and collapse. On the other hand, the oscillatory mode exist because of shear layer interaction with the corner of after-body. As, it  \textcolor{black}{is understood} that increase in Mach number at same altitude increases the incoming momentum of the flow, which can cause the bow shock envelop squeeze or  \textcolor{black}{become} closer to body. This may lead to different interaction of bow shock, shear layer  \textcolor{black}{and re-circulation zone} with corner. 
\begin{figure}[!htb]
\centering
\includegraphics[scale= 0.75]{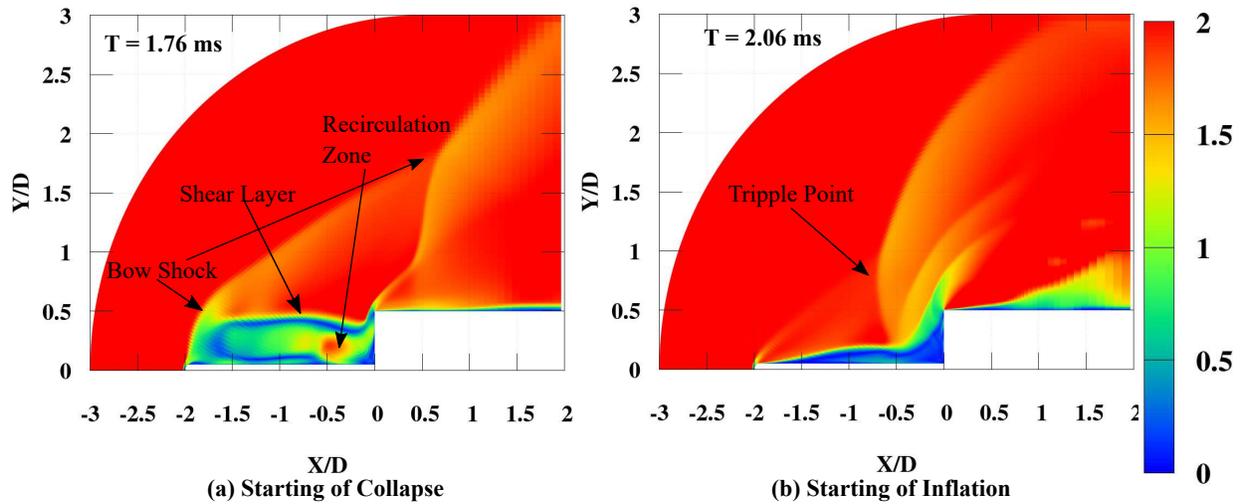}
\caption{\label{Fig04} Mach Contours at different time steps for free-stream Mach number 2.0}
\end{figure}
Figure \ref{Fig04} shows the Mach contours at two extreme time-steps of a cycle at Mach 2. \textcolor{black}{Also, it highlights in general various flow features responsible for pulsation and oscillation mode, while interacting with the corner of after-body. In this case at Mach 2, the} flow behaviour exhibits the large amplitude pulsation mode, because of available large range of bow-shock movement due to expanded bow-shock envelop at low Mach number. The starting of collapse or end of inflation is time-stepped at $t = 1.76$ ms. as shown in Fig. \ref{Fig04}a . At this time step, the frontal bow shock remains farthest away in front of the after body and it interacts with the after body shock at far downstream location. The region behind the bow shock remain subsonic, however inner region near the frontal face travels towards the incoming flow at supersonic speed. At the end of inflation phase, the incoming mass in front of flat face, starts returning backward and escapes from the corner edge. The end of collapse or starting of inflation is shown at time step $t = 2.06$ ms in Fig. \ref{Fig04}b The region between spike tip and front face squeezed to smallest. The bow shock interacts with the front face shock above the spike. The backward shock becomes closer to front face, which may reflect in pressure plot Fig. \ref{Fig02}b as sudden increase in pressure. The envelop of backward shock is bigger because of low Mach number of incoming flow. After the end of collapse, the region between the front face and spike tip starts inflating and the pulsation cycle continues.
\begin{figure}[!htb]
\centering
\includegraphics[scale= 0.75]{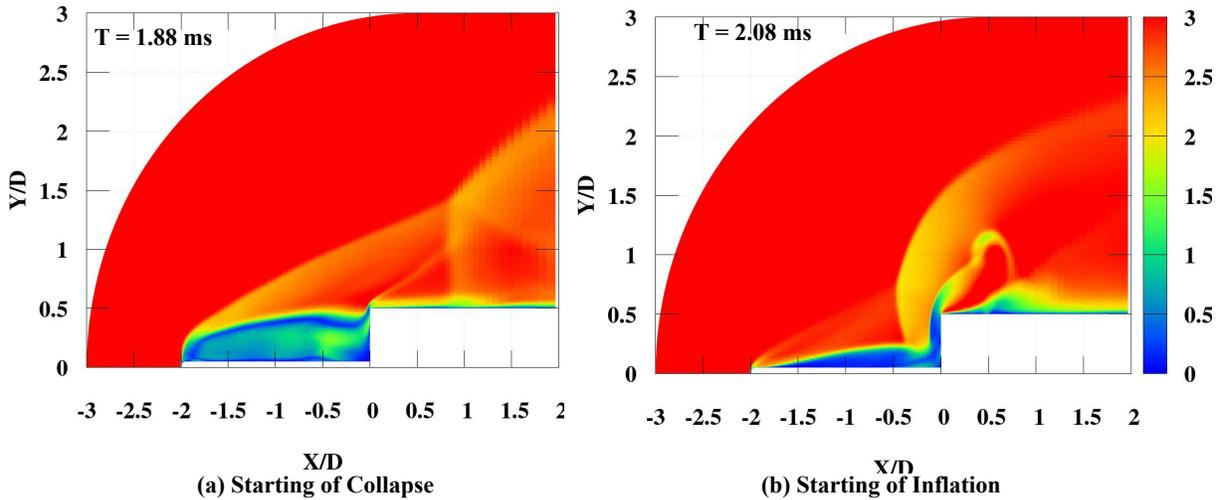}
\caption{\label{Fig05} Mach Contours at different time steps for free-stream Mach number 3.0}
\end{figure}
Figure \ref{Fig05} shows the two extreme movement of bow shock in front of spike blunt nose at Mach 3. At Mach 3, similar shock fluctuations in pulsation mode can be seen. However, the envelop of shock waves at Mach 3 is slightly squeezed at both time steps. The interaction of front bow shock to corner or backward shock at lower Y-location and farther downstream location for Mach 3 in comparison to Mach 2 as seen in Fig. \ref{Fig05}a. The end of collapse phase at Mach 3 shows the higher pressure zone closer to the flat face. It causes higher pressure increase in Fig. \ref{Fig02}b for Mach 3 case than compare to Mach 2 case. The time duration between shown time-steps are lower than Mach 2 case, reflects higher frequency pulsation at Mach 3.
\begin{figure}[!htb]
\centering
\includegraphics[scale= 0.75]{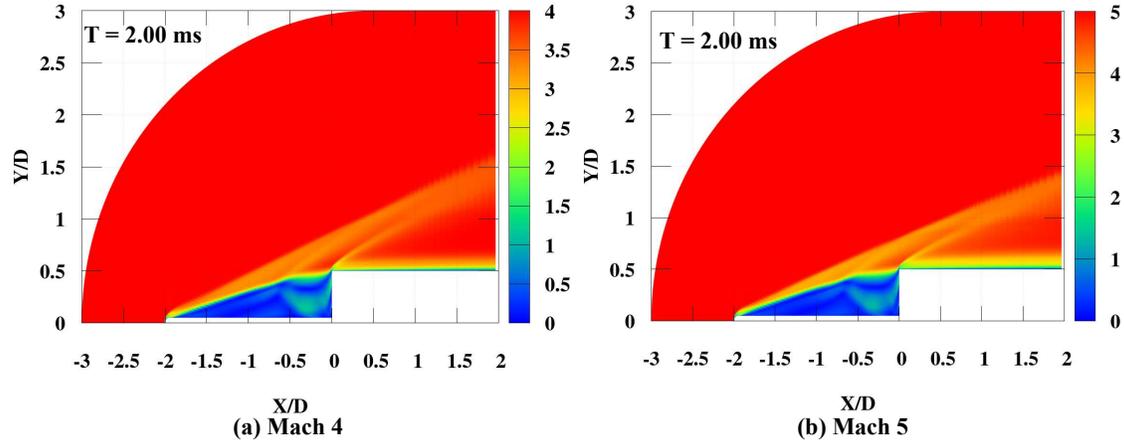}
\caption{\label{Fig06} Mach Contours for almost stable flow field at free-stream Mach number 4.0 and 5.0}
\end{figure}
Further, Fig. \ref{Fig06} shows the relatively stable bow shock cases at Mach 4 and 5. At these two Mach numbers the bow shock envelope remains above the corner and shear layer behind it interacts with the corner, which leads to very small amplitude oscillations in case of Mach 4 and slightly higher amplitude oscillation in case of Mach 5 as seen in Fig. \ref{Fig02}b and \ref{Fig03}b. At these Mach numbers the large amplitude pulsation below Mach 4 has been completely turned into small amplitude oscillation mode. However, the reason of this change in unsteady behaviour is the shock positioning slightly above the corner, which may depend not only on Mach number, but also several parameters such as spike length and after-body cross-section area.
\begin{figure}[!htb]
\centering
\includegraphics[scale= 0.75]{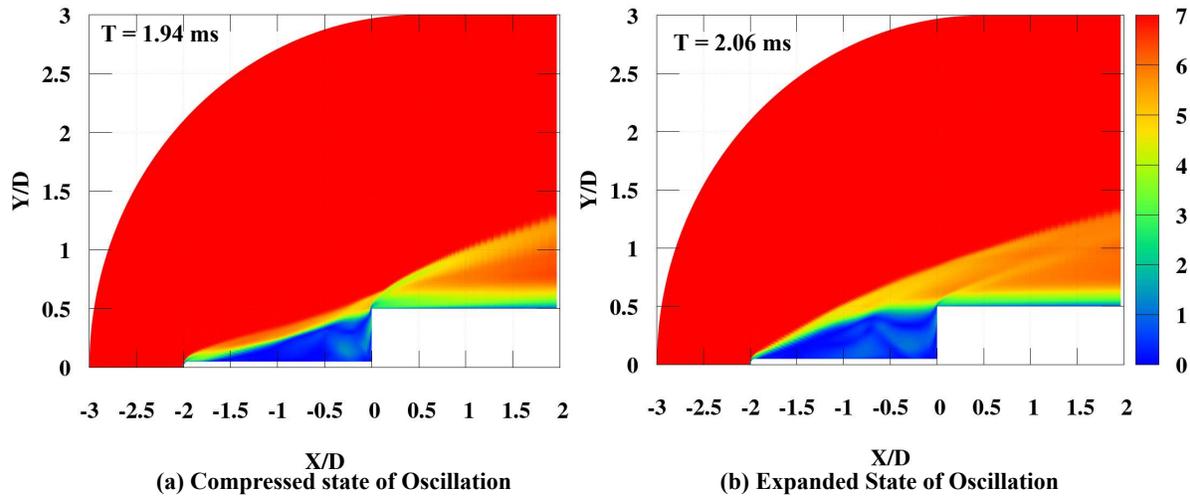}
\caption{\label{Fig07} Mach Contours for oscillatory mode at free-stream Mach number 7.0}
\end{figure}
Figure \ref{Fig07} shows the Mach contours for oscillation mode of bow shock in front of spiked blunt nose at Mach 7. The behaviour of Mach 6 and Mach 7 are same, only difference is that the bow shock oscillates at higher frequency and higher amplitude than at Mach 6 as evident from Fig. 3. The bow shock envelop remains in the vicinity of after-body corner and energetic shear layer interacts with the corner in the manner of breathe in and breathe-out causes the shape of bow shock changes from concave to convex. The change in bow shock shape causes the oscillatory mode, which doesn't have large amplitude as compare to pulsation mode in Fig \ref{Fig02}. The amplitude of change in pressure and drag coefficient increases as the Mach number increases and remains highest for Mach number 7 among the studied cases. If the current simulation at Mach 6 are compared with similar computation performed at Mach 6 \cite{shashank1} for L/D =2 hemispherical tipped spike, where the free-stream pressure and temperature, 253 Pa and 60 K, respectively. The flow-field at the mentioned free-stream condition exhibited the pulsation mode, opposite to what was observed here. Hence, it can be said that the bow-shock unsteadiness is however dependent on Mach number at same free-stream conditions, while there are other governing parameters, which allow shock to position in the manner to exhibit pulsation and oscillation modes.

\subsection{\textcolor{black}{Unsteady Flow} Analysis}
\begin{figure}[!htb]
\centering
\includegraphics[scale= 0.74]{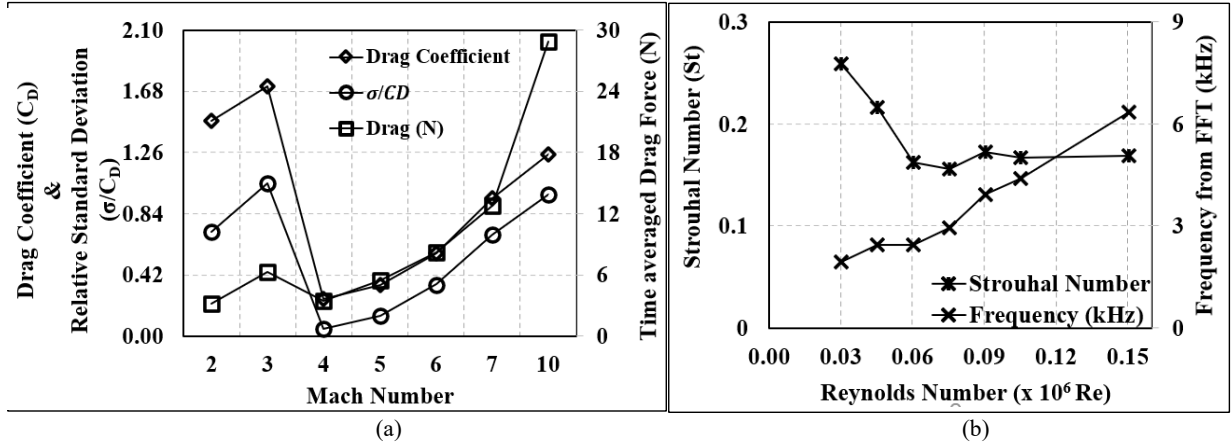}
\caption{\label{Fig08} (a) Drag Coefficient \& Relative Standard Deviation along with absolute Drag Force (on Secondary Axis) (b) Strouhal Number along with computed Frequency (kHz) from FFT of Pressure data}
\end{figure}
\textcolor{black}{In this section, various parameters are compared to understand and characterize the pulsation and oscillation modes for all the computed cases. One more data point at Mach 10 is also added here to have better discussion. The time-averaged drag coefficient (for the time 1 ms to 4 ms), as well as relative standard deviation are plotted in Fig. \ref{Fig08}a. The relative standard deviation can be defined as ratio of calculated standard deviation ($\sigma$) and mean value, which can represent the level of fluctuations in time-series signal. Further, the drag-coefficient may be sightly misleading in understanding the effect of Mach number, hence the absolute values of time-averaged drag forces are plotted on secondary axis, varying with Mach number. The time-averaged drag force increases with increase in Mach number, except slight decrease between Mach 3 and 4, where the bow shock instability transit from pulsation mode to oscillation mode. The time-averaged drag coefficient also shows increase between Mach 2 and 3, while it shows sharp decrease between Mach 3 to 4, when the transition between pulsation and oscillation mode occur. The relative standard deviation (RSD) plot ($\sigma/ C_D$ or $\sigma/Drag$) provides understanding of the amplitude of fluctuations with respect to the mean drag coefficient or drag force. At Mach 2, the RSD plot shows 70 \% of mean, which increases above 100 \% of mean at Mach 3 in pulsation mode. While the transition between pulsation to oscillation occur at Mach 4, the RSD drops to 5\% of mean and further increases with in increase in Mach number from 5 \% of mean to 97 \% of mean drag at Mach 10. Further, the pressure plots as shown in Fig. \ref{Fig02} and \ref{Fig03} are used to compute the Fast Fourier Transformation for all the computed cases. The extracted frequency from FFT has been plotted along with Strouhal Number in Fig. \ref{Fig08}b}. The Strouhal number can be computed from the following relation:
\begin{equation}
St = \frac{fL}{V_{\infty}}
\end{equation}
where, $f$ is the frequency of fluctuations, L is characteristic length, spike length in this case and $V_{\infty}$ is free-stream velocity. 
\textcolor{black}{The frequency and Strouhal Numbers are plotted with respect to Reynolds number at each Mach number based on cross-section diameter of the after-body here. It is evident from the previous section results that the frequency of unsteadiness increases with increase in Mach number, whereas the amplitude increases or decreases with a minimum 1.95 kHz at Mach 2 to a maximum of 6.34 kHz at Mach 10. The computed Strouhal number shows interesting trends. It decreases with change in Mach numbers from 2 to 4, while the flow behavior changes from pulsation to oscillation mode. Further, the Strouhal number remain almost constant and below 0.2 through out the Mach number change from 4 to 10, while the bow-shock oscillation frequency and amplitude increases. The Strouhal number computed by Vashishtha et al. \cite{shashank1} was found to be at 0.25 in free-stream pressure and temperature conditions 253 Pa and 60 K, respectively at Mach 6 flow and spike length of $L/D = 2$. The bow shock in previous study fluctuates in pulsation mode with frequency of 2.93 kHz. Hence, it can be said that the flow behaviour is not only affected by Mach number, but also incoming momentum which governs the shock positioning with respect to after-body corner Also, it may be said here that the higher Strouhal number values (more than 0.2 in this case) reflects the pulsation mode, while lower values shows the oscillation mode. However the limit of Strouhal number exhibiting pulsation or oscillation mode may be dependent on spike length along with other flow parameter.}




\section{Conclusions}
This study focuses on unsteady behaviour of bow shock wave in front of spiked blunt nose at various Mach numbers ranging from supersonic Mach 2 to hypersonic Mach 7 along with addition Mach number 10. In this study, the Mach number is changed by changing the speed, while keeping the free-stream conditions same. It was found that the drag coefficient as well as front wall pressure fluctuates in highly unsteady manner with couple of peaks and troughs at low Mach number of Mach 2 and 3, while the flow transition to stable with very small amplitude oscillations at Mach 4. At higher Mach number 5, 6 and 7, the bow shock only exhibits the oscillation mode with higher frequency. The amplitude of oscillation increases with increase in Mach number leads to higher pressure at front face. While comparing the same Mach number and spike length at different free-stream pressure and temperature of previous study, it was found that pulsation or oscillation mode depends on bow shock positioning irrespective of Mach number. While Mach number is one of the important parameter affecting flow unsteadiness at same free-stream condition, however free-stream condition, spike length and other parameters may affect the shock positioning with respect to after-body corner and may lead to different pulsation and oscillation modes. \textcolor{black}{The Strouhal number limit for characterizing the pulsation or oscillation mode with flat after-body and spike length of $L/D =2$ is observed as 0.2 for Mach number variation between 2 to 10}.


\nocite{*}
\bibliography{iCraft2021}

\providecommand{\noopsort}[1]{}\providecommand{\singleletter}[1]{#1}%
\begin{thebibliography}{15}%
\makeatletter
\providecommand \@ifxundefined [1]{%
 \@ifx{#1\undefined}
}%
\providecommand \@ifnum [1]{%
 \ifnum #1\expandafter \@firstoftwo
 \else \expandafter \@secondoftwo
 \fi
}%
\providecommand \@ifx [1]{%
 \ifx #1\expandafter \@firstoftwo
 \else \expandafter \@secondoftwo
 \fi
}%
\providecommand \natexlab [1]{#1}%
\providecommand \enquote  [1]{``#1''}%
\providecommand \bibnamefont  [1]{#1}%
\providecommand \bibfnamefont [1]{#1}%
\providecommand \citenamefont [1]{#1}%
\providecommand \href@noop [0]{\@secondoftwo}%
\providecommand \href [0]{\begingroup \@sanitize@url \@href}%
\providecommand \@href[1]{\@@startlink{#1}\@@href}%
\providecommand \@@href[1]{\endgroup#1\@@endlink}%
\providecommand \@sanitize@url [0]{\catcode `\\12\catcode `\$12\catcode
  `\&12\catcode `\#12\catcode `\^12\catcode `\_12\catcode `\%12\relax}%
\providecommand \@@startlink[1]{}%
\providecommand \@@endlink[0]{}%
\providecommand \url  [0]{\begingroup\@sanitize@url \@url }%
\providecommand \@url [1]{\endgroup\@href {#1}{\urlprefix }}%
\providecommand \urlprefix  [0]{URL }%
\providecommand \Eprint [0]{\href }%
\providecommand \doibase [0]{http://dx.doi.org/}%
\providecommand \selectlanguage [0]{\@gobble}%
\providecommand \bibinfo  [0]{\@secondoftwo}%
\providecommand \bibfield  [0]{\@secondoftwo}%
\providecommand \translation [1]{[#1]}%
\providecommand \BibitemOpen [0]{}%
\providecommand \bibitemStop [0]{}%
\providecommand \bibitemNoStop [0]{.\EOS\space}%
\providecommand \EOS [0]{\spacefactor3000\relax}%
\providecommand \BibitemShut  [1]{\csname bibitem#1\endcsname}%
\let\auto@bib@innerbib\@empty
\bibitem [{\citenamefont {Ahmed}\ and\ \citenamefont {Qin}(2020)}]{ahmed2020}%
  \BibitemOpen
  \bibfield  {author} {\bibinfo {author} {\bibfnamefont {M.~Y.}\ \bibnamefont
  {Ahmed}}\ and\ \bibinfo {author} {\bibfnamefont {N.}~\bibnamefont {Qin}},\
  }\bibfield  {title} {\enquote {\bibinfo {title} {Forebody shock control
  devices for drag and aero-heating reduction: A comprehensive survey with a
  practical perspective},}\ }\href {\doibase 10.1016/j.paerosci.2019.100585}
  {\bibfield  {journal} {\bibinfo  {journal} {Progress in Aerospace Sciences}\
  }\textbf {\bibinfo {volume} {112}},\ \bibinfo {pages} {100585} (\bibinfo
  {year} {2020})}\BibitemShut {NoStop}%
\bibitem [{\citenamefont {Maull}(1960)}]{maull}%
  \BibitemOpen
  \bibfield  {author} {\bibinfo {author} {\bibfnamefont {D.~J.}\ \bibnamefont
  {Maull}},\ }\bibfield  {title} {\enquote {\bibinfo {title} {Hypersonic flow
  over axially symmetric spiked bodies},}\ }\href {\doibase
  10.1017/S0022112060000815} {\bibfield  {journal} {\bibinfo  {journal}
  {Journal of Fluid Mechanics}\ }\textbf {\bibinfo {volume} {8}},\ \bibinfo
  {pages} {584–592} (\bibinfo {year} {1960})}\BibitemShut {NoStop}%
\bibitem [{\citenamefont {Wood}(1961)}]{wood}%
  \BibitemOpen
  \bibfield  {author} {\bibinfo {author} {\bibfnamefont {C.~J.}\ \bibnamefont
  {Wood}},\ }\bibfield  {title} {\enquote {\bibinfo {title} {Hypersonic flow
  over spiked cones},}\ }\href {\doibase 10.1017/S0022112062000427} {\bibfield
  {journal} {\bibinfo  {journal} {Journal of Fluid Mechanics}\ }\textbf
  {\bibinfo {volume} {12}},\ \bibinfo {pages} {614--624} (\bibinfo {year}
  {1961})}\BibitemShut {NoStop}%
\bibitem [{\citenamefont {Kenworthy}(1978)}]{Kenworthy}%
  \BibitemOpen
  \bibfield  {author} {\bibinfo {author} {\bibfnamefont {M.~A.}\ \bibnamefont
  {Kenworthy}},\ }\emph {\bibinfo {title} {A study of unstable axisymmetric
  separation in high speed flows}},\ \href {http://hdl.handle.net/10919/76093}
  {\bibinfo {type} {Ph.d. thesis}},\ \bibinfo  {school} {Virginia Polytechnic
  Institute and State University} (\bibinfo {year} {1978})\BibitemShut
  {NoStop}%
\bibitem [{\citenamefont {Vashishtha}, \citenamefont {Watanabe},\ and\
  \citenamefont {Suzuki}(2015{\natexlab{a}})}]{utokyo1}%
  \BibitemOpen
  \bibfield  {author} {\bibinfo {author} {\bibfnamefont {A.}~\bibnamefont
  {Vashishtha}}, \bibinfo {author} {\bibfnamefont {Y.}~\bibnamefont
  {Watanabe}}, \ and\ \bibinfo {author} {\bibfnamefont {K.}~\bibnamefont
  {Suzuki}},\ }\bibfield  {title} {\enquote {\bibinfo {title} {Study of shock
  shape in front of concave, convex and flat arc in hypersonic flow},}\ }\href
  {https://ci.nii.ac.jp/naid/120006827780/en/} {\bibfield  {journal} {\bibinfo
  {journal} {JAXA-SP-14-010}\ ,\ \bibinfo {pages} {127--132}} (\bibinfo {year}
  {2015}{\natexlab{a}})}\BibitemShut {NoStop}%
\bibitem [{\citenamefont {Vashishtha}, \citenamefont {Watanabe},\ and\
  \citenamefont {Suzuki}(2015{\natexlab{b}})}]{utokyo2}%
  \BibitemOpen
  \bibfield  {author} {\bibinfo {author} {\bibfnamefont {A.}~\bibnamefont
  {Vashishtha}}, \bibinfo {author} {\bibfnamefont {Y.}~\bibnamefont
  {Watanabe}}, \ and\ \bibinfo {author} {\bibfnamefont {K.}~\bibnamefont
  {Suzuki}},\ }\enquote {\bibinfo {title} {Study of bow-shock instabilities in
  front of hemispherical shell at hypersonic \textsc{M}ach number 7},}\ in\
  \href {\doibase 10.2514/6.2015-2638} {\emph {\bibinfo {booktitle} {45th AIAA
  Fluid Dynamics Conference AIAA 2015-2638}}}\ (\bibinfo {year}
  {2015})\BibitemShut {NoStop}%
\bibitem [{\citenamefont {Vashishtha}, \citenamefont {Watanabe},\ and\
  \citenamefont {Suzuki}(2016)}]{utokyo3}%
  \BibitemOpen
  \bibfield  {author} {\bibinfo {author} {\bibfnamefont {A.}~\bibnamefont
  {Vashishtha}}, \bibinfo {author} {\bibfnamefont {Y.}~\bibnamefont
  {Watanabe}}, \ and\ \bibinfo {author} {\bibfnamefont {K.}~\bibnamefont
  {Suzuki}},\ }\bibfield  {title} {\enquote {\bibinfo {title} {Bow-shock
  instability and its control in front of hemispherical concave shell at
  hypersonic \textsc{M}ach number 7},}\ }\href {\doibase
  10.2322/tastj.14.Pe_121} {\bibfield  {journal} {\bibinfo  {journal}
  {Transaction of the JSASS, Aerospace Technology Japan}\ }\textbf {\bibinfo
  {volume} {14}},\ \bibinfo {pages} {121--128} (\bibinfo {year}
  {2016})}\BibitemShut {NoStop}%
\bibitem [{\citenamefont {Vashishtha}, \citenamefont {Callaghan},\ and\
  \citenamefont {Nolan}(2021)}]{vashishtha2021}%
  \BibitemOpen
  \bibfield  {author} {\bibinfo {author} {\bibfnamefont {A.}~\bibnamefont
  {Vashishtha}}, \bibinfo {author} {\bibfnamefont {D.}~\bibnamefont
  {Callaghan}}, \ and\ \bibinfo {author} {\bibfnamefont {C.}~\bibnamefont
  {Nolan}},\ }\bibfield  {title} {\enquote {\bibinfo {title} {Drag control by
  hydrogen injection in shocked stagnation zone of blunt nose},}\ }\href
  {\doibase 10.1088/1757-899x/1024/1/012110} {\bibfield  {journal} {\bibinfo
  {journal} {IOP Conference Series: Materials Science and Engineering}\
  }\textbf {\bibinfo {volume} {1024}},\ \bibinfo {pages} {012110} (\bibinfo
  {year} {2021})}\BibitemShut {NoStop}%
\bibitem [{\citenamefont {Feszty}\ \emph {et~al.}(2000)\citenamefont {Feszty},
  \citenamefont {Richards}, \citenamefont {Badcock},\ and\ \citenamefont
  {Woodgate}}]{festzy1}%
  \BibitemOpen
  \bibfield  {author} {\bibinfo {author} {\bibfnamefont {D.}~\bibnamefont
  {Feszty}}, \bibinfo {author} {\bibfnamefont {B.}~\bibnamefont {Richards}},
  \bibinfo {author} {\bibfnamefont {K.}~\bibnamefont {Badcock}}, \ and\
  \bibinfo {author} {\bibfnamefont {M.}~\bibnamefont {Woodgate}},\ }\bibfield
  {title} {\enquote {\bibinfo {title} {Numerical simulation of a pulsating flow
  arising over an axisymmetric spiked blunt body at \textsc{M}ach 2.21 and
  \textsc{M}ach 6.00},}\ }\href {\doibase 10.1007/s001930000065} {\bibfield
  {journal} {\bibinfo  {journal} {Shock Waves}\ }\textbf {\bibinfo {volume}
  {10}},\ \bibinfo {pages} {323--331} (\bibinfo {year} {2000})}\BibitemShut
  {NoStop}%
\bibitem [{\citenamefont {Feszty}, \citenamefont {Badcock},\ and\ \citenamefont
  {Richards}(2004{\natexlab{a}})}]{festzy2}%
  \BibitemOpen
  \bibfield  {author} {\bibinfo {author} {\bibfnamefont {D.}~\bibnamefont
  {Feszty}}, \bibinfo {author} {\bibfnamefont {K.~J.}\ \bibnamefont {Badcock}},
  \ and\ \bibinfo {author} {\bibfnamefont {B.~E.}\ \bibnamefont {Richards}},\
  }\bibfield  {title} {\enquote {\bibinfo {title} {Driving mechanisms of
  high-speed unsteady spiked body flows, part 1: Pulsation mode},}\ }\href
  {\doibase 10.2514/1.9034} {\bibfield  {journal} {\bibinfo  {journal} {AIAA
  Journal}\ }\textbf {\bibinfo {volume} {42}},\ \bibinfo {pages} {95--106}
  (\bibinfo {year} {2004}{\natexlab{a}})}\BibitemShut {NoStop}%
\bibitem [{\citenamefont {Feszty}, \citenamefont {Badcock},\ and\ \citenamefont
  {Richards}(2004{\natexlab{b}})}]{festzy3}%
  \BibitemOpen
  \bibfield  {author} {\bibinfo {author} {\bibfnamefont {D.}~\bibnamefont
  {Feszty}}, \bibinfo {author} {\bibfnamefont {K.~J.}\ \bibnamefont {Badcock}},
  \ and\ \bibinfo {author} {\bibfnamefont {B.~E.}\ \bibnamefont {Richards}},\
  }\bibfield  {title} {\enquote {\bibinfo {title} {Driving mechanism of
  high-speed unsteady spiked body flows, part 2: Oscillation mode},}\ }\href
  {\doibase 10.2514/1.9035} {\bibfield  {journal} {\bibinfo  {journal} {AIAA
  Journal}\ }\textbf {\bibinfo {volume} {42}},\ \bibinfo {pages} {107--113}
  (\bibinfo {year} {2004}{\natexlab{b}})}\BibitemShut {NoStop}%
\bibitem [{\citenamefont {Vashishtha}\ and\ \citenamefont
  {Khurana}(2021{\natexlab{a}})}]{shashank1}%
  \BibitemOpen
  \bibfield  {author} {\bibinfo {author} {\bibfnamefont {A.}~\bibnamefont
  {Vashishtha}}\ and\ \bibinfo {author} {\bibfnamefont {S.}~\bibnamefont
  {Khurana}},\ }\bibfield  {title} {\enquote {\bibinfo {title} {On unsteady
  flow analysis of a round spike blunt nose afterbody in \textsc{M}ach 6
  flow},}\ }\href {\doibase 10.1088/1757-899x/1024/1/012017} {\bibfield
  {journal} {\bibinfo  {journal} {IOP Conference Series: Materials Science and
  Engineering}\ }\textbf {\bibinfo {volume} {1024}},\ \bibinfo {pages} {012017}
  (\bibinfo {year} {2021}{\natexlab{a}})}\BibitemShut {NoStop}%
\bibitem [{\citenamefont {Vashishtha}\ and\ \citenamefont
  {Khurana}(2021{\natexlab{b}})}]{shashank2}%
  \BibitemOpen
  \bibfield  {author} {\bibinfo {author} {\bibfnamefont {A.}~\bibnamefont
  {Vashishtha}}\ and\ \bibinfo {author} {\bibfnamefont {S.}~\bibnamefont
  {Khurana}},\ }\enquote {\bibinfo {title} {Pulsating flow investigation for
  spiked blunt-nose body in hypersonic flow and its control},}\ in\ \href
  {\doibase 10.2514/6.2021-0839} {\emph {\bibinfo {booktitle} {AIAA Scitech
  Forum, AIAA-2021-0839}}}\ (\bibinfo {year} {2021})\BibitemShut {NoStop}%
\bibitem [{\citenamefont {Liou}(2006)}]{liou}%
  \BibitemOpen
  \bibfield  {author} {\bibinfo {author} {\bibfnamefont {M.-S.}\ \bibnamefont
  {Liou}},\ }\bibfield  {title} {\enquote {\bibinfo {title} {A sequel to
  \textsc{AUSM}, \textsc{P}art \textsc{II}: \textsc{AUSM+-}up for all
  speeds},}\ }\href {\doibase 10.1016/j.jcp.2005.09.020} {\bibfield  {journal}
  {\bibinfo  {journal} {J. of Computational Physics}\ }\textbf {\bibinfo
  {volume} {214}},\ \bibinfo {pages} {137–170} (\bibinfo {year}
  {2006})}\BibitemShut {NoStop}%
\bibitem [{\citenamefont {Shu}\ and\ \citenamefont {Osher}(1988)}]{shu}%
  \BibitemOpen
  \bibfield  {author} {\bibinfo {author} {\bibfnamefont {C.-W.}\ \bibnamefont
  {Shu}}\ and\ \bibinfo {author} {\bibfnamefont {S.}~\bibnamefont {Osher}},\
  }\bibfield  {title} {\enquote {\bibinfo {title} {Efficient implementation of
  essentially non-oscillatory shock-capturing schemes},}\ }\href {\doibase
  10.1016/0021-9991(88)90177-5} {\bibfield  {journal} {\bibinfo  {journal} {J.
  of Computational Physics}\ }\textbf {\bibinfo {volume} {77}},\ \bibinfo
  {pages} {439--471} (\bibinfo {year} {1988})}\BibitemShut {NoStop}%
\end{thebibliography}%
\end{document}